# Tricritical Behavior of Two-Dimensional Scalar Field Theories

M. Asorey [*]

*Instituto de Física. Universidade Federal do Rio de Janeiro*

*Cidade Universitaria. Ilha do fundão, Cx. P. 68528 21945-970, Rio de Janeiro, RJ, Brazil*

J.G. Esteve, F. Falceto

*Departamento de Física Teórica. Facultad de Ciencias*

*Universidad de Zaragoza. 50009 Zaragoza. Spain*

J. Salas

*Department of Physics. New York University*

*4 Washington Place. New York, NY 10003. USA*

## Abstract

We compute by Monte Carlo numerical simulations the critical exponents of two-dimensional scalar field theories at the $\lambda\phi^6$ tricritical point. The results are in agreement with the Zamolodchikov conjecture based on conformal invariance.

Typeset using REVTEX

---

[*]Permanent address: Departamento de Física Teórica. Facultad de Ciencias. Universidad de Zaragoza. 50009 Zaragoza. Spain

In the last decade there has been a considerable progress in the classification of universality classes of two-dimensional (2D) critical phenomena in terms of conformal field theories [1]. For most of the universality classes there are integrable models whose critical exponents can be exactly calculated. However, the critical behavior of other 2D models has to be investigated by other means because they cannot be exactly solved. Self–interacting scalar theories with even polynomial interaction

$$S(\phi) = \frac{1}{2}\sum_{i=1}^{2}(\partial_i\phi)^2 + \sum_{r=1}^{N}\lambda_{2r}\phi^{2r} \qquad (1)$$

belong to this second class of models.

The phase structure of these theories is very complicated. It contains in general $l$-critical manifolds ($l = 2, \ldots, N$) of dimensionality $d_l = N+1-l$ in the space of the coupling constants $\{\lambda_{2r}\}_{r=1}^{r=N}$. A.B. Zamolodchikov [2] conjectured that the leading multicritical behavior of these models for arbitrary $N \geq 2$ corresponds to the conformal models of the unitary series with central charge

$$c = 1 - \frac{6}{(N+1)(N+2)}. \qquad (2)$$

However, there is no direct proof of this conjecture in terms of explicit calculations of correlation functions or finite–size scaling (FSS) analysis of thermodynamical quantities.

The critical exponents can be obtained from the natural identification pointed out by Zamolodchikov between relevant local perturbations and primary fields $\phi_{p,q}$ of the corresponding conformal theory. These fields possess scaling dimensions given by the Kac formula

$$h_{p,q} = \frac{[(N+2)p - (N+1)q]^2 - 1}{4(N+1)(N+2)}; \qquad (3)$$
$$1 \leq p \leq N; \quad 1 \leq q \leq N+1$$

The order parameter $\phi$ (i.e. the magnetization) is identified with the most relevant non–trivial primary field $\phi \approx \phi_{2,2}$ with scaling dimension $h_{2,2} = 3/[4(N+1)(N+2)]$. The critical exponent $\eta$ can be read off from the decay of the two–point correlation function

$$\langle \phi(x)\phi(0)\rangle \approx \frac{1}{|x|^{d-2+\eta}}, \tag{4}$$

at large distances $|x| \gg 1$. Thus, $\eta = 4h_{2,2}$ for $d = 2$.

The energy density operator $\epsilon$ is given by the renormalized composite field $:\phi^2:$ and it is identified with the primary field $\epsilon \approx \phi_{3,3}$ (resp. $\phi_{1,3}$) for $N \geq 3$ (resp. $N = 2$). The corresponding scaling dimension is $h_{3,3} = 2/[(N+1)(N+2)]$ (resp. $h_{1,3} = 1/2$). The two–point correlator of the energy density has a similar behavior to that of Eq. (4); but now the power is related to the critical exponent $\nu$ [3]

$$\langle \epsilon(x)\epsilon(0)\rangle \approx \frac{1}{|x|^{2(d-1/\nu)}}, \tag{5}$$

This implies that $1/\nu = 2(1 - h_{3,3})$ (resp. $1/\nu = 1$). The remaining exponents can be obtained via the well–known scaling relations [4]. The result is displayed in Table I. Notice that the Gaussian critical exponents are recovered in the limit $N \to \infty$.

The free scalar bosonic field ($N = 1$) is the only model of the family which is exactly solvable. The critical massless regime describes a conformal field theory with central charge $c = 1$. Thus, its critical exponents are given by the case $N = \infty$ in Table I.

The first non–trivial case corresponds to $N = 2$. This model contains one critical line in the coupling constant space with the Gaussian model at one end ($\lambda_4 = \lambda_2 = 0$) and the Ising model ($\lambda_4 \to \infty, \lambda_2 \to -\infty, \lambda_4/\lambda_2 \to$ constant) at the other one [5]. It is generally believed that the whole critical line (except the Gaussian endpoint) belongs to the Ising universality class ($c = 1/2$). However, some authors have claimed that this could not be the case [6], although their results are not very conclusive and they might be interpreted as strong corrections to scaling, as suggested in [7].

The case $N = 3$ is the first one where tricritical points are found. It is remarkable the striking difference between the conjectured value for $\eta = 3/20$ and the prediction coming from standard perturbation theory $\eta = 1/500$ [4]. This model was also considered by Hamber [8], who used an analysis based on an approximate renormalization–group transformation on the equivalent one–dimensional quantum Hamiltonian. A tricritical point was found and the leading eigenvalue was equal to $y_H = 1/\nu = 1.661$ (resp. 1.817) for block size $b = 2$ (resp. $b = 3$). From Table I we see that Zamolodchikov's prediction is $1/\nu = 1.8$, very close to the latter result. However, a complete study of this model is still lacking.

In this letter the validity of this conjecture is analyzed by explicit calculation of the corresponding critical exponents in the case $N = 3$. We consider a square $L \times L$ lattice with unit spacing and periodic boundary conditions. To each site $i$ of the lattice we assign a continuous unbounded spin $\phi_i$. The partition function $\mathcal{Z}$ of this model on the lattice is given in terms of the discretized Euclidean action $S$

$$\mathcal{Z} = \int \mathcal{D}\phi \; \mathrm{e}^{-S(\phi)}; \; S(\phi) = - \sum_{<i,j>} \phi_i \phi_j + \sum_i V(\phi_i) \tag{6}$$

where $\mathcal{D}\phi = \prod_i d\phi_i$; and the first sum is over all the nearest–neighbor pairs of spins. The one–particle potential $V$ is defined as

$$V(\phi) = \frac{1}{2}(4 - m^2)\phi^2 + \sum_{r=2}^{N=3} \lambda_{2r} \phi^{2r} \tag{7}$$

For convenience, we have identified $\lambda_2$ with minus the mass squared $\lambda_2 = -m^2$.

The phase diagram of this model can be described qualitatively by the mean field approximation [4]. For any $\lambda_6 > 0$ there is a curve of singular points in the $(m^2, \lambda_4)$ subspace. For positive $\lambda_4$, the transition is second–order. However, when $\lambda_4$ becomes negative, there is a point where the transition changes its character to first order. This is the tricritical point. Actually, there is a line of tricritical points in the whole coupling constant space $(m^2, \lambda_4, \lambda_6)$. It is expected that all these tricritical points belong to the same universality class, irrespective of the value of $\lambda_6 > 0$. The numerical results reported here correspond to the choice $\lambda_6 = 1$.

We have carefully analyzed the phase diagram by means of the variational method introduced in [9]. It turned out to be very accurate for the search of critical points in the subspace $(m^2, \lambda_4, \lambda_6 = 1)$. This method also allows to distinguish clearly among critical, tricritical and first–order points. In practice, two different critical points has been selected.

One is an ordinary (Ising–like) critical point and the other one is a critical point in the vicinity of the tricritical point. The region around those points has been explored by Monte Carlo (MC) simulations, which confirmed that they are actually very close to the real critical points.

To simulate the model given by Eq. (6) we have used a heat–bath algorithm based on one proposed in ref. [5] for the case $N = 2$. This one has the important drawback that it only works for $m^2 < 4$. We have generalized their algorithm for arbitrary $N$ by adding and subtracting a $m_0^2 \phi^2$ term to the potential in Eq. (7). The probability density of choosing the new spins can be written as in [5]: it is a product of a Gaussian distribution (with their $m^2$ substituted by $m^2 - 2m_0^2$) and an exponential term $h(\phi) = \exp(m_0^2 \phi^2 - \lambda_4 \phi^4 - \lambda_6 \phi^6)$, which is bounded if $\lambda_6 > 0$. This latter function can be normalized by adding some constant to the potential (7) which is irrelevant when computing expectation values. The parameter $m_0^2$ is free, except for the constraint $2m_0^2 > m^2 - 4$. Thus, it can be tuned to maximize the performance of the algorithm (see below).

The following quantities have been measured: energy density $E = \frac{1}{L^2}\langle S \rangle$, specific heat $C_v = \frac{1}{L^2}\sigma^2(S)$ (where $\sigma(\cdot)$ is the standard deviation), magnetization $M = \frac{1}{L^2}\langle |\sum_i \phi_i| \rangle$, magnetic susceptibility $\chi = \frac{1}{L^2}\sigma^2(|\sum_i \phi_i|)$, the mean value of $\phi^2$ (i.e. $M' = \frac{1}{L^2}\langle \sum_i \phi_i^2 \rangle$) and the corresponding "susceptibility" $\chi' = \frac{1}{L^2}\sigma^2(\sum_i \phi_i^2)$.

Given the values of $\lambda_4$ and $\lambda_6 = 1$ (selected by the variational method), for each $L$ we vary $m^2$ until a maximum in any of the quantities $Q = \{C_v, \chi, \chi'\}$ is attained. This is done in practice by using the spectral density method (SDM) [10]. For each quantity $Q$ we obtain the position $m_L^2(Q)$ of the peak and the height of such peak $Q_L^{max}$. First, some preliminary runs were made to select the region of interest in the $m^2$ axis. The longest runs were carried out at masses such that their distance to the predicted position of the peaks are less than the range of validity of the SDM $\Delta m^2 \approx 2/[L\sqrt{\chi'}\,]$ [5]. In some occasions several masses were selected to cover the three peaks. A summary of the runs is contained in Table II.

Using standard FSS analysis [11] we can extract from the MC data the critical mass

$m_c^2$ and the critical exponent associated with each quantity $Q$ (i.e. the exponent $\rho$ defined by $Q(m^2) \sim |m^2 - m_c^2|^{-\rho}$) divided by $\nu$. In our case, we can estimate the ratios $\gamma/\nu$, $\alpha/\nu$ and $\gamma'/\nu$ (this one corresponds to the susceptibility $\chi'$). Notice that Zamolodchikov's conjecture implies that $\gamma' = \alpha$. Finally, it is also worthy to consider the quantity $M(m_L^2(\chi))$ which is defined as the value of the magnetization at the point where the magnetic susceptibility peaks. This quantity is expected to go to zero as $L^{-\beta/\nu}$ for $L$ large enough [12]. The ratio $\beta/\nu$ is related directly to $\gamma/\nu$ via the scaling relation

$$\beta/\nu = 1 - \gamma/(2\nu) \tag{8}$$

Thus, the validity of this important relation could be tested numerically in a very clean way for this model.

To check the quality of our results we have estimated the autocorrelation time for the magnetization ($\tau_M$) and the energy density using the methods of ref. [13]. It turned out that the slowest mode was the magnetization, although we do not exclude the existence of another slow mode not considered here. If we fit our results to $\tau_M = AL^z$, we obtain a dynamical critical exponent roughly equal to $z \approx 2.0$ for the two MC simulations reported here (see Table II). This is in agreement with the conclusions of [14]. We have also found that the constant $A$ is larger for the simulation near the tricritical point. Unfortunately, due to the large autocorrelation time of the heat–bath algorithm, we have been unable to obtain very good statistics. For the largest lattice we only could do $\sim 10^3 \ \tau_M$ measures (resp. $\sim 2 \cdot 10^3 \ \tau_M$ measures) for the MC simulation close to the tricritical point (resp. Ising–like critical point). For the smaller lattices these numbers are one order of magnitude larger. Furthermore, the possibility of performing a reasonable simulation on larger lattices is beyond of our computer facilities. Finally, to save on CPU time, we have skipped a certain number of MC steps between measures. This number varies from 20 in the smaller lattices to $2 \cdot 10^3$ in the largest one (see Table II).

First, we have analyzed a critical point on the expected $\lambda\phi^4$ regime by fixing $\lambda_4 = 0$ (and $\lambda_6 = 1$). The results of the simulations are displayed in Table III. The three extrapolated critical values of $m_c^2$ are compatible and converge to a value $m_c^2 \approx 4.67 \pm 0.08$. The results for the ratios of critical exponents are the following

$$\frac{\gamma}{\nu} = 1.73 \pm 0.02 \qquad \frac{\beta}{\nu} = 0.14 \pm 0.02 \qquad \frac{\alpha}{\nu} = \frac{\gamma'}{\nu} = 0. \qquad (9)$$

The null results have been obtained from the failure of any power fit (e.g. the best fit gives $\alpha/\nu \sim 0.1$ with a very large value of $\chi^2 \sim 6$ with one degree of freedom). On the other hand, the logarithmic fit is reasonably good ($\chi^2 \sim 0.5$ with one degree of freedom). Similar numbers are obtained for the susceptibility $\chi'$. Errors in (9) have been estimated as the double of the standard deviation (i.e. 95% confidence level). The results agree within errors with Zamolodchikov's conjectures for the $\lambda \phi^4$ ($N = 2$) critical point ($\gamma/\nu = 1.75, \beta/\nu = 0.125, \alpha/\nu = \gamma'/\nu = 0$). We also remark that the scaling relation Eq. (8) is satisfied within statistical errors.

We now analyze the results for the second critical point, of coordinates $\lambda_4 = -4.55$ and $\lambda_6 = 1$ (see Table IV). This value of $\lambda_4$ was selected by the variational method mentioned above. When the mass $m^2$ is chosen suitably, it is presumably very close to the real tricritical point $(m_{tc}^2, \lambda_{4,tc}, \lambda_{6,tc} = 1)$. In such a case, even if we are not exactly at the tricritical point, we can extract information about the critical exponents of the tricritical regime. For not too large lattice size $L$, the behavior near the tricritical point is expected to be very similar to that *at* the tricritical point (see for instance [15]). That is because for finite $L$ any thermodynamic quantity is an analytical function of the coupling constants, even at the critical points. Close to the tricritical point the behavior of thermodynamic quantities interpolate between the tricritical regime and the infinite volume asymptotic regime. If we are really very close to the tricritical point the size of the lattice where the crossover between the two regimes appears might be very large. However, if the size of the system becomes large enough, the true asymptotic behavior is always recovered. The window where the tricritical regime is observed increases as we approach the tricritical point, therefore, it is very important to make the simulation very close to it.

The three extrapolated critical values of $m_{tc}^2$ are compatible and converge to a value

$m_{tc}^2 \approx -9.815 \pm 0.003$. The ratios of critical exponents obtained from the FSS analysis

$$\frac{\gamma}{\nu} = 1.81 \pm 0.02; \qquad \frac{\beta}{\nu} = 0.083 \pm 0.005$$
$$\frac{\alpha}{\nu} = \frac{\gamma'}{\nu} = 1.54 \pm 0.02 \qquad (10)$$

are in fairly agreement with the values predicted according to Zamolodchikov's conjecture, $\gamma/\nu = 1.85, \beta/\nu = 0.077, \alpha/\nu = \gamma'/\nu = 1.60$.

The exponents (10) are very different of those associated to the $\lambda\phi^4$ critical behavior (specially the ratios $\alpha/\nu = \gamma'/\nu$ and $\beta/\nu$) and of those characteristic of a first–order phase transition (i.e. $\gamma/\nu = \alpha/\nu = \gamma'/\nu = 2$, $\beta/\nu = 0$ [16]). However, all of them except $\beta/\nu$ are slightly below the conjectured results and the differences are several standard deviations and thus, they are statistically significant. This fact can be understood if our point is a little bit away from the true tricritical point, on the line of Ising–like critical points (i.e. $\lambda_4 > \lambda_{4,tc}$) and the lattices sizes are not too large. On the other hand, the value of $\beta/\nu$ is slightly higher than the predicted one. This is in agreement with the fact that the scaling relation (8) is also verified. In this context, it would be very interesting to study in greater detail the crossover from Ising–like behavior to first–order transitions passing through the tricritical point.

In summary, our results confirm the validity of Zamolodchikov conjecture for the case $N = 3$, and in other words, the power of conformal field theory methods into the description of the critical behavior of 2D field theories. Our analysis makes more plausible that the conjecture holds for the whole series of $\lambda\phi^{2n}$ multicritical points, although a rigorous proof is still lacking.

## ACKNOWLEDGEMENTS

J. S. thanks the members of Departamento de Física Teórica of Universidad de Zaragoza for their hospitality during the completion of this work. We also acknowledge CNPq and CICYT for partial financial support (grants AEN93-219, AEN94-218).

TABLES

TABLE I. Critical exponents predicted by conformal invariance according to Zamolodchikov's conjecture.

| Exponents | $N = 2$ | $N = 3$ | $N \geq 3$ | $N = \infty$ |
|---|---|---|---|---|
| $\nu$ | 1 | $\frac{5}{9}$ | $\frac{1}{2} + \frac{1}{(N+1)(N+2)-2}$ | $\frac{1}{2}$ |
| $\gamma$ | $\frac{7}{4}$ | $\frac{37}{36}$ | $\frac{2(N+1)(N+2)-3}{2(N+1)(N+2)-4}$ | 1 |
| $\alpha$ | 0 | $\frac{8}{9}$ | $\frac{(N+1)(N+2)-4}{(N+1)(N+2)-2}$ | 1 |
| $\eta$ | $\frac{1}{4}$ | $\frac{3}{20}$ | $\frac{3}{(N+1)(N+2)}$ | 0 |
| $\beta$ | $\frac{1}{8}$ | $\frac{1}{24}$ | $\frac{3}{4(N+1)(N+2)-8}$ | 0 |
| $\delta$ | 15 | $\frac{77}{3}$ | $\frac{4(N+1)(N+2)}{3} - 1$ | $\infty$ |

TABLE II. Monte Carlo simulations performed in the coupling constant space $(m^2, \lambda_4, \lambda_6 = 1)$. We also display the size of the lattice and the total number of measures and iterations.

| $\lambda_4$ | $L$ | $m^2$ | # measures | # iterations |
| --- | --- | --- | --- | --- |
| 0 | 8 | 4.20 | $10^5$ | $2 \cdot 10^6$ |
| 0 | 8 | 4.70 | $10^5$ | $2 \cdot 10^6$ |
| 0 | 16 | 4.40 | $10^5$ | $2 \cdot 10^6$ |
| 0 | 16 | 4.60 | $10^5$ | $2 \cdot 10^6$ |
| 0 | 32 | 4.50 | $4 \cdot 10^4$ | $2 \cdot 10^6$ |
| 0 | 32 | 4.60 | $4 \cdot 10^4$ | $2 \cdot 10^6$ |
| 0 | 64 | 4.55 | $4 \cdot 10^4$ | $4 \cdot 10^6$ |
| 0 | 64 | 4.60 | $4 \cdot 10^4$ | $4 \cdot 10^6$ |
| -4.55 | 8 | -9.80 | $4 \cdot 10^4$ | $2 \cdot 10^6$ |
| -4.55 | 16 | -9.81 | $10^4$ | $2 \cdot 10^6$ |
| -4.55 | 16 | -9.82 | $10^4$ | $2 \cdot 10^6$ |
| -4.55 | 32 | -9.816 | $1.05 \cdot 10^4$ | $1.05 \cdot 10^7$ |
| -4.55 | 64 | -9.815 | $1.45 \cdot 10^4$ | $2.9 \cdot 10^7$ |

TABLE III. For the simulation performed at $\lambda_4 = 0$ and $\lambda_6 = 1$, we show the values of the maximum of the specific heat and its position $(C_{v,L}^{max}, m_L^2(C_v))$ as functions of the lattice size $L$. The same values for the magnetic susceptibility $(\chi_L^{max}, m_L^2(\chi))$ and the $\chi'$ susceptibility $(\chi_L'^{max}, m_L^2(\chi'))$ are also displayed, as well as the magnetization $M(m_L^2(\chi))$ at the pseudo–critical point $m_L^2(\chi)$. The parentheses denote the errors (one standard deviation, $\sim 67\%$ confidence level) in the last relevant figures.

| $L$ | 8 | 16 | 32 | 64 |
|---|---|---|---|---|
| $C_{v,L}^{max}$ | 1.7637(97) | 2.228(13) | 2.675(31) | 3.120(30) |
| $m_L^2(C_v)$ | 4.6857(83) | 4.6079(60) | 4.5952(77) | 4.6176(37) |
| $\chi_L^{max}$ | 2.291(10) | 7.458(51) | 24.95(32) | 82.3(12) |
| $m_L^2(\chi)$ | 4.1844(58) | 4.3817(35) | 4.5039(36) | 4.5712(27) |
| $\chi_L'^{max}$ | 0.3684(18) | 0.4327(25) | 0.4927(49) | 0.5462(47) |
| $m_L^2(\chi')$ | 4.512(13) | 4.5576(93) | 4.5863(87) | 4.6094(47) |
| $M(m_L^2(\chi))$ | 0.4183(12) | 0.37205(86) | 0.3382(22) | 0.3059(22) |

TABLE IV. The same as Table III but for the values $\lambda_4 = -4.55$ and $\lambda_6 = 1$.

| $L$ | 8 | 16 | 32 | 64 |
|---|---|---|---|---|
| $C_{v,L}^{max}$ | 4.368(24) | 9.62(12) | 25.18(26) | 69.77(48) |
| $m_L^2(C_v)$ | −9.80742(65) | −9.81543(76) | −9.81518(16) | −9.81485(5) |
| $\chi_L^{max}$ | 16.603(71) | 59.78(39) | 212.9(16) | 739.2(51) |
| $m_L^2(\chi)$ | −9.82953(84) | −9.81843(76) | −9.81577(17) | −9.81495(6) |
| $\chi_L'^{max}$ | 27.78(10) | 83.4(71) | 247.8(20) | 717.1(52) |
| $m_L^2(\chi')$ | −9.82751(83) | −9.81800(75) | −9.81565(17) | −9.81492(5) |
| $M(m_L^2(\chi))$ | 0.69989(85) | 0.6555(16) | 0.6176(12) | 0.5842(15) |